\documentclass[conference]{IEEEtran}
\IEEEoverridecommandlockouts

\usepackage[dvipdfmx]{graphicx}
\usepackage{url}

\usepackage{multirow}

\usepackage{cite}
\usepackage{amsmath,amssymb,amsfonts}
\usepackage{algorithmic}
\usepackage{graphicx}
\usepackage{xcolor}

\usepackage{multirow}
\usepackage{colortbl}

\def\BibTeX{{\rm B\kern-.05em{\sc i\kern-.025em b}\kern-.08em
 T\kern-.1667em\lower.7ex\hbox{E}\kern-.125emX}}
\begin{document}

\title{Is an investor stolen their profits by mimic investors? Investigated by an agent-based model
\thanks{Note that the opinions contained herein are solely those of the authors and do not necessarily reflect those of SPARX Asset Management Co., Ltd.}
}

\author{
\IEEEauthorblockN{Takanobu Mizuta}
\IEEEauthorblockA{\textit{SPARX Asset Management Co. Ltd.} \\ Tokyo, Japan \\ https://orcid.org/0000-0003-4329-0645}
\and
\IEEEauthorblockN{Isao Yagi}
\IEEEauthorblockA{\textit{Faculty of Informatics} \\
\textit{Kogakuin University}\\
Tokyo, Japan \\ https://orcid.org/0000-0003-0119-1366}
}

\maketitle

\IEEEpubidadjcol

\begin{abstract}

Some investors say increasing investors with the same strategy decreasing their profits per an investor. On the other hand, some investors using technical analysis used to use same strategy and parameters with other investors, and say that it is better. Those argues are conflicted each other because one argues using with same strategy decreases profits but another argues it increase profits. However, those arguments have not been investigated yet. In this study, the agent-based artificial financial market model(ABAFMM) was built by adding ``additional agents''(AAs) that includes additional fundamental agents (AFAs) and additional technical agents (ATAs) to the prior model. The AFAs(ATAs) trade obeying simple fundamental(technical) strategy having only the one parameter. We investigated earnings of AAs when AAs increased. We found that in the case with increasing AFAs, market prices are made stable that leads to decrease their profits. In the case with increasing ATAs, market prices are made unstable that leads to gain their profits more.

\end{abstract}


\section{Introduction} \label{s1}

Some investors say increasing investors with the same strategy decreasing their profits per an investor. On the other hand, some investors using technical analysis used to use same strategy and parameters with other investors, and say that it is better. Those argues are conflicted each other because one argues using with same strategy decreases profits but another argues it increase profits. However, those arguments have not been investigated yet. 

As so many factors affect price formation in actual markets, empirical studies cannot be conducted to isolate the direct effect of increasing investors using exactly same investment strategy. In contrast, an artificial market model, which is an agent-based artificial financial market model(ABAFMM)\cite{mizuta2025mono} can isolate the pure contributions of increasing such investors. ABAFMMs have greatly contributed to design a financial market that works well, and that includes making and/or modulating detailed regulations and/or rules\cite{mizuta2025mono}. The design of a financial market is very important for the development and maintenance of an advanced economy, but designing it is not easy because changes in detailed rules, even those that seem trivial, sometimes have unexpectedly large impacts and side effects in a financial market, which is a complex system. Traditional economics cannot treat a financial market as a complex system in which micro–macro interaction and feedback loops have played essential roles, because traditional economics can only treat macrophenomena and micro processes separately. ABAFMM can do it, however.

In this study, the ABAFMM was built by adding ``additional agents''(AAs) that includes additional fundamental agents (AFAs) and additional technical agents (ATAs) to the prior model of Mizuta and Yagi \cite{mizuta2025mono}. The AFAs(ATAs) trade obeying simple fundamental(technical) strategy having only the one parameter. We investigated earnings of AAs when AAs increased.

\begin{figure}[t] 
\begin{center}
\includegraphics[scale=0.40]{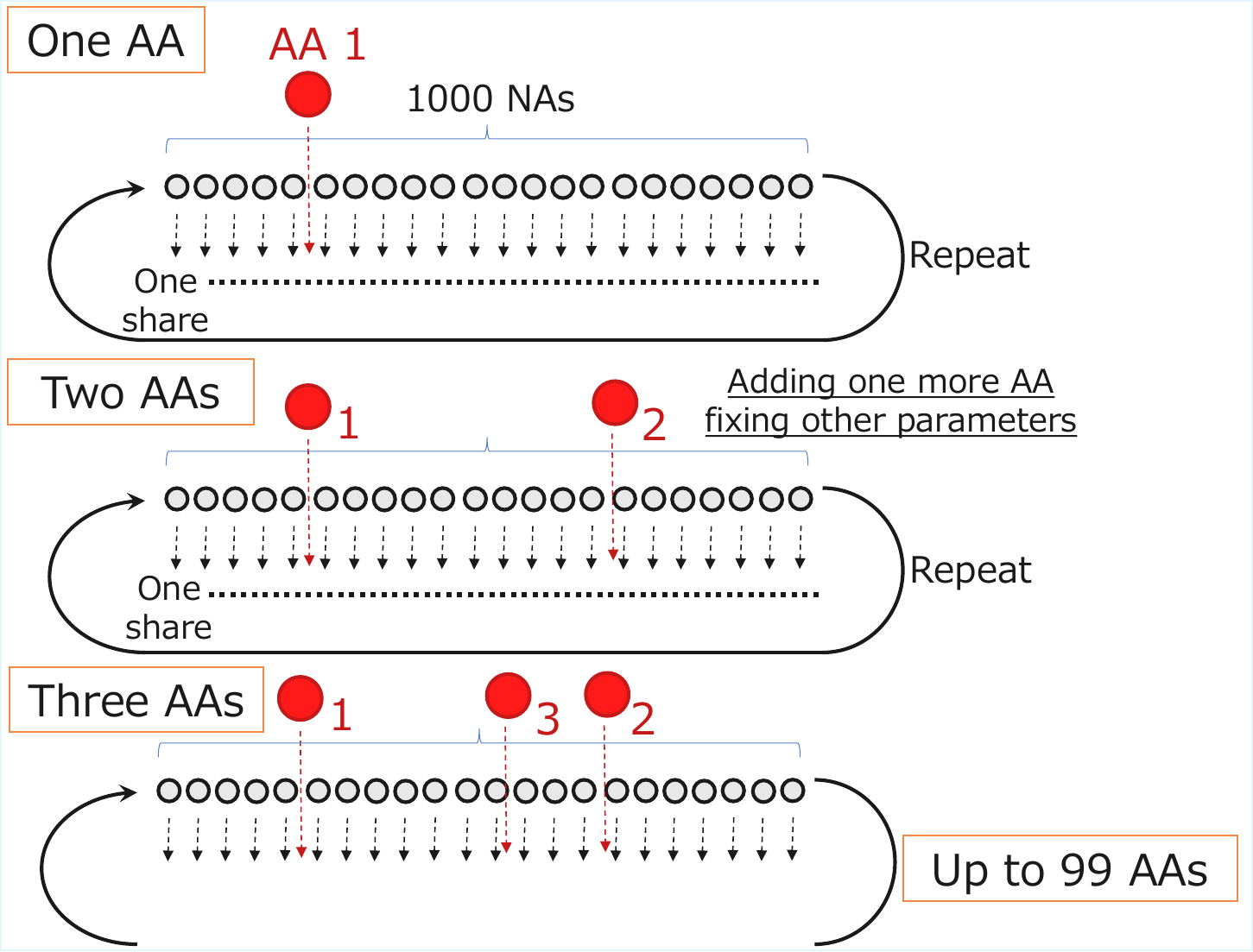}
\end{center}
\caption{Timing of AAs placing orders.}
\label{p01}
\end{figure}

\section{Model} \label{s2}

An exchange uses a continuous double auction to determine the market price. In the auction mechanism, multiple buyers and sellers compete to buy and sell stocks in the market, and transactions can occur at any time whenever an offer to buy and one to sell match. The minimum unit of a price change is $\delta P$. The buy-order and sell-order prices are rounder down and up to the nearest fraction, respectively.

\subsection{Normal agents(NAs)} \label{s2.1}

To replicate the nature of price formation in actual financial markets, we introduced the normal agents(NAs) to model a general investor. The number of NAs is $n$ for the each exchange. The NAs can short sell freely. The holding positions are not limited, so the NAs can take an infinite number of shares for both long and short positions.  Time $t$ increases time $t$ one when an NA place an order. 

NAs always order only one share. First, NA $1$ places an order to buy or sell a stock; then NA $2$ places an order to buy or sell. After that, NAs $3,4,,,n$ each place orders to buy or sell. After the NA $n$ placed an order, going back to the first NA, $1$ places an order to buy or sell, and at NAs $2,3,,,,n$ each place orders to buy or sell, and this cycle is repeated.

An NA determines the order price and buys or sells using a combination of fundamental and technical analysis strategies to form an expectation of the stock return. The expected return of NA $j$ at $t$ is
\begin{equation}
r^{t}_{e,j} = (w_{1,j} \ln{\frac{P_f}{P^{t-1}}} + w_{2,j}\ln{\frac{P^{t-1}}{P^{t-\tau _ j-1}}}+w_{3,j} \epsilon ^t _j )/\Sigma_i^3 w_{i,j}, \label{eq1}
\end{equation}
where $w_{i,j}$ is the weight of term $i$ for NA $j$ and is independently determined by random variables uniformly distributed on the interval $(0,w_{i,max})$ at the start of the simulation for each NA. $\ln$ is the natural logarithm. $P_f$ is a fundamental value and is a constant. $P^t$ is a mid-price (the average of the highest buy-order price and the lowest sell-order price) at $t$, and $\epsilon ^t _ j$ is determined by random variables from a normal distribution with average $0$ and variance $\sigma _ \epsilon$ at $t$. $\tau_j$ is independently determined by random variables uniformly distributed on the interval $(1,\tau _{max})$ at the start of the simulation for each NA\footnote{When $t< \tau _ j$, the second term of Eq. (\ref{eq1}) is zero.}.

The first term in Eq. (\ref{eq1}) represents a fundamental strategy: the NA expects a positive return when the market price is lower than the fundamental value, and vice versa. The second term represents a technical analysis strategy using a historical return: the NA expects a positive return when the historical market return is positive, and vice versa. The third term represents noise.

After the expected return has been determined, the expected price is
\begin{equation}
P^t_{e,j}= P^{t-1} \exp{(r^t_{e,j})}.
\end{equation}

Order prices are scattered around the expected price $P^t_{e,j}$ to replicate many waiting limit orders. An order price $P^t_{o,j}$ is 
\begin{equation}
P^t_{o,j}=P^t_{e,j}+P_d(2\rho ^t _j-1),
\end{equation}
where $\rho ^t_j$ is determined by random variables uniformly distributed on the interval $(0,1)$ at $t$ and $P_d$ is a constant. This means that $P^t_{o,j}$ is determined by random variables uniformly distributed on the interval $(P^t_{e,j}-P_d, P^t_{e,j}+P_d)$ 

Whether the NA buys or sells is determined by the magnitude relationship between $P^t_{e,j}$ and $P^t_{o,j}$. Here\footnote{When $t<t_c$, to generate enough waiting orders, the NA places an order to buy one share when $P_f>P^t_{o,j}$, or to sell one share when $P_f<P^t_{o,j}$. \label{ft01}},

\begin{itemize}
    \item[] when $P^t_{e,j}>P^t_{o,j}$, the NA places an buy order, 
    \item[] when $P^t_{e,j}<P^t_{o,j}$, the NA places an sell order. 
\end{itemize}

The remaining order is canceled $t_c$.

\subsection{Additional Agents (AAs)} \label{s2.2}

Additional agents(AAs) that includes additional fundamental agents (AFAs) and additional technical agents (ATAs). The AFAs(ATAs) trade obeying simple fundamental(technical) strategy having only the one parameter. We investigated earnings of AAs when AAs increased. Fig \ref{p01} illustrated when AAs place orders. When the one AA is added it places an order at some random time. It places orders at the fixed same time after second loops. When the two AAs are added first added AA places an order at the fixed same time and the second added AA places an order at some random time. They place orders at the fixed same time after second loops. When more than three AAs are added, they place orders as the same way. AA is added one by one, final number of AAs is $n_a$.

\subsubsection{Additional Fundamental Agents (AFAs)} \label{s2.2.2}

Here, we denote that the highest buy-order price is $P^t_b$ and the lowest sell-order price is $P^t_s$ at $t$.

When $P^t_s<P_f$, AFAs buy one share when they have no shares and buy two shares when they short-sold one share. In this case, they finally hold one share. They do not place orders when they already have one share.

When $P^t_b>P_f$, AFAs short-sell one share when they have no shares and short-sell two shares(sell one share and short-sell one share) when they short-sell one share one share. In this case, they finally short-sell one share. They do not place orders when they already short-sold one share.

In other cases, they do not place orders.

\subsubsection{Additional Technical Agents (ATAs)} \label{s2.2.3}

Here, all ATAs have same the one parameter, $ta$ that how long do they refer the price ago.

When $P^t_s>P^{t-ta}$, ATAs buy one share when they have no shares and buy two shares when they short-sold one share. In this case, they finally hold one share. They do not place orders when they already have one share.

When $P^t_b<P^{t-ta}$, ATAs short-sell one share when they have no shares and short-sell two shares(sell one share and short-sell one share) when they short-sell one share one share. In this case, they finally short-sell one share. They do not place orders when they already short-sold one share.

In other cases, they do not place orders.

\begin{figure}[t] 
\begin{center}
\includegraphics[scale=0.40]{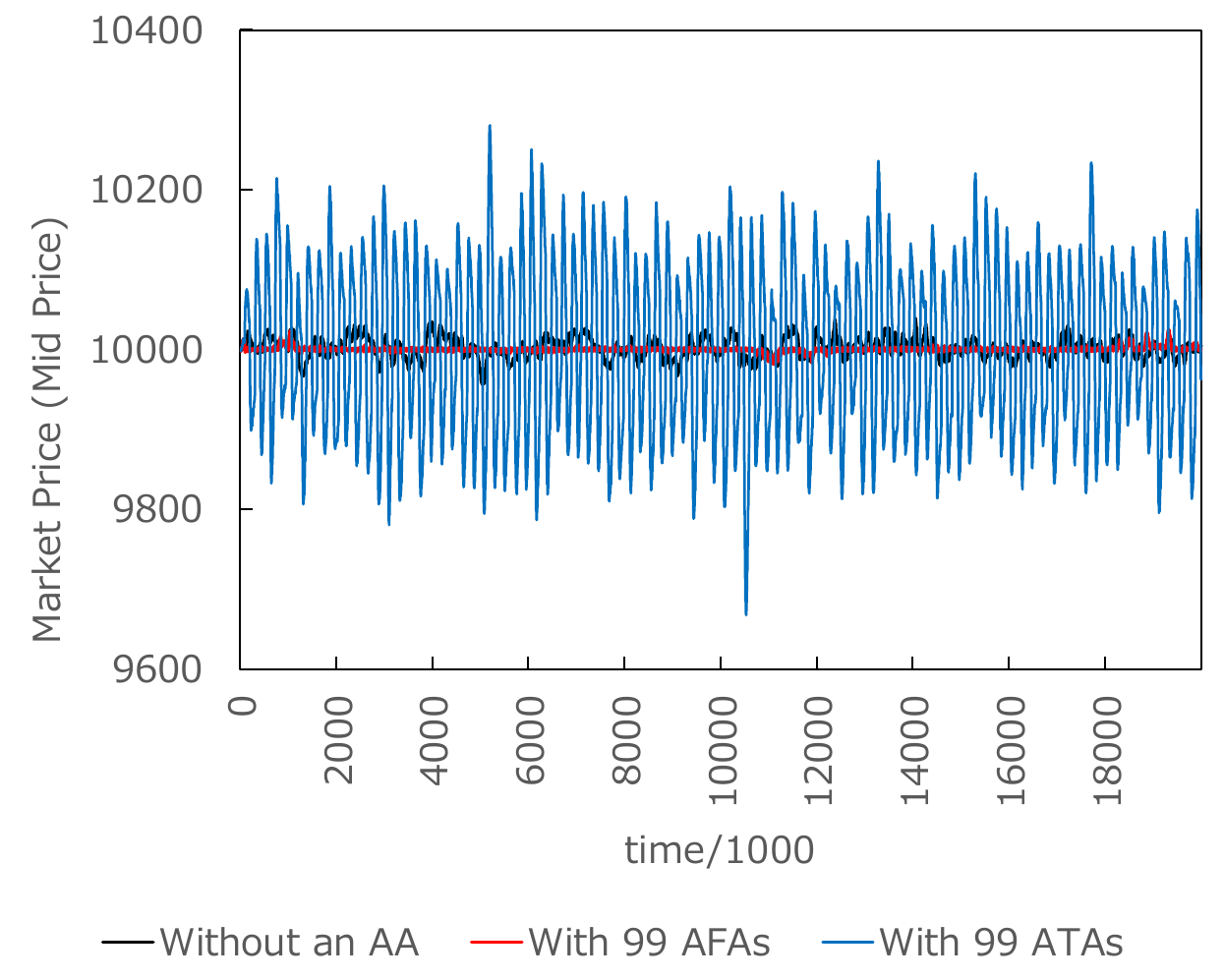}
\end{center}
\caption{Time evolution of market prices $P^t$ without an AA, with 99 AFAs and 99 ATAs.}
\label{e01}
\end{figure}

\begin{figure}[t] 
\begin{center}
\includegraphics[scale=0.40]{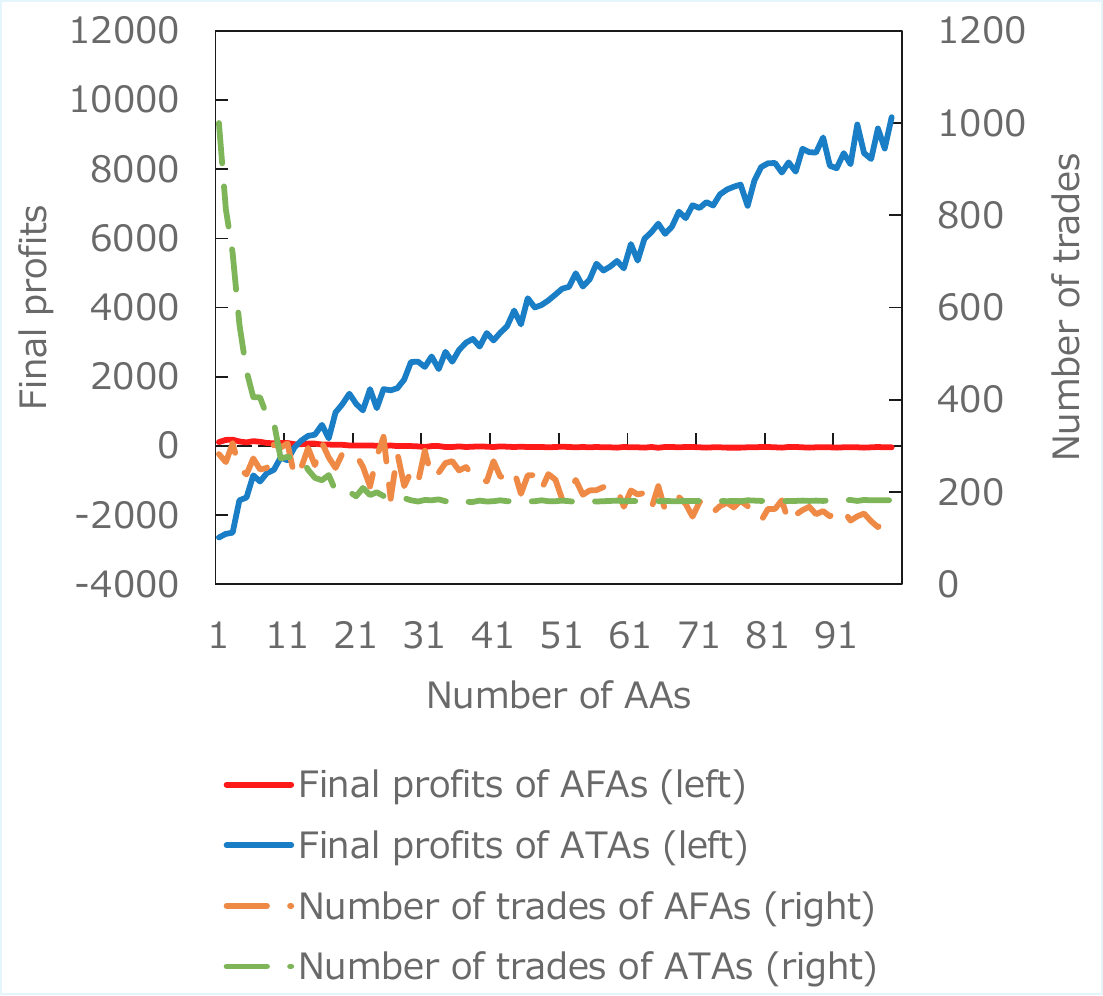}
\end{center}
\caption{The average of final profits and number of trades of AAs for $n_a$.}
\label{e03}
\end{figure}

\begin{figure}[t] 
\begin{center}
\includegraphics[scale=0.25]{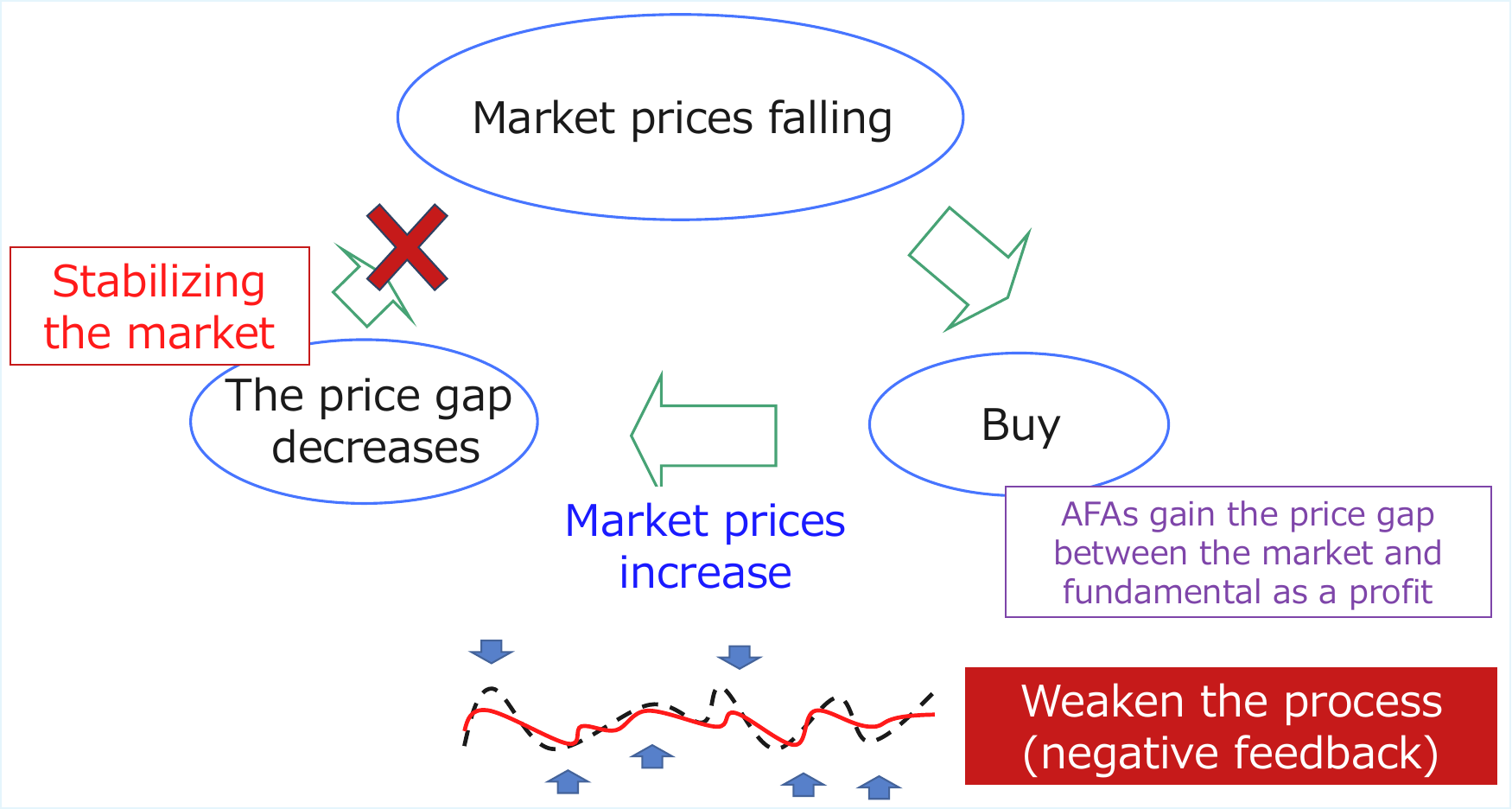}
\end{center}
\caption{Mechanism of AFAs stabilizing markets.}
\label{p02}
\end{figure}

\begin{figure}[t] 
\begin{center}
\includegraphics[scale=0.25]{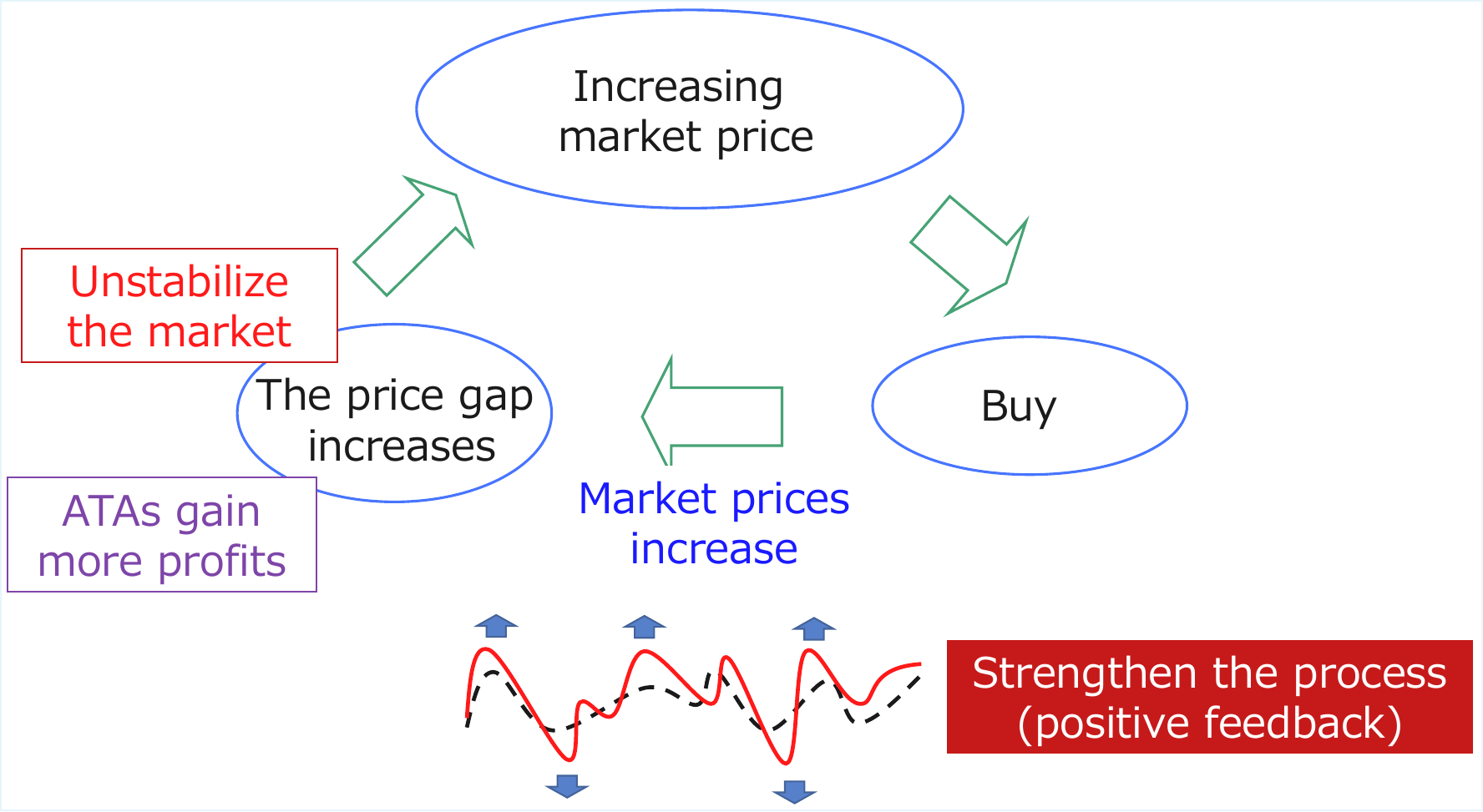}
\end{center}
\caption{Mechanism of ATAs unstabilize markets.}
\label{p03}
\end{figure}

\section{Simulation results}
\label{s4}

In this study, I set $\delta P=0.01, P_{f}=10000$, and for the NAs, $n=1000, w_{1,max}=1, w_{2,max}=100, w_{3,max}=1, \tau _ {max}=10000, \sigma _ \epsilon = 0.03, P_d= 1000$, and $t_c=10000$ as same as Mizuta et. al. \cite{mizuta2025mono}. For AAs, I set $na=99, ta=100000$. The simulations ran to $t=t_e=20000000$. 

Fig. \ref{e01} shows time evolution of market prices $P^t$ without AAs, with 99 AFAs and 99 AFAs. The market is stabilized with AFAs. This means that increasing AFAs make the market more stable. On the other hand, the market is very unstable with ATAs. ATAs amplify the market variation. 

Fig. \ref{e03} shows the average of final profits and number of trades of AAs for $n_a$. Holding shares at the simulation finished were evaluated with $P_f$. In the case with ATAs, more ATAs, more profits. Number of trades was rapidly declined in $n_a<20$ but it was stable in $n_a>20$. This shows that ATAs gain more profits to amplify the variation of market prices by impacts of themselves' trades. On the other hand, in the case with AFAs the market prices were stables.

Fig. \ref{p02} illustrated mechanism of AFAs stabilizing markets. Here, we discuss a falling case. When market prices fell, AFAs buy. These buys leads market prices to increase and converge with the fundamental price. AFAs gain the price gap between the market and fundamental as a profit, but the gap is decreasing by themselves' trades that leads market stable. So, trades of AFAs make the price gap decreased and weaken this profit process. This process is a negative feedback process. 

Fig. \ref{p03} illustrated mechanism of ATAs unstabilize markets. Here, we discuss an increasing market price case. When market prices increase, ATAs buy. These buys leads market prices to increase more. Increasing market prices more by themselves' trades leads more profits and market unstable. So, trades of ATAs make the market prices unstable more, gain more profits and strengthen this profit process. This process is a positive feedback process.

\section{Summary}

In this study, the ABAFMM was built by adding AAs that includes AFAs and ATAs to the prior model of Mizuta and Yagi \cite{mizuta2025mono}. The AFAs(ATAs) trade obeying simple fundamental(technical) strategy having only the one parameter. We investigated earnings of AAs when AAs increased. We found that in the case with increasing AFAs, market prices are made stable that leads to decrease their profits. In the case with increasing ATAs, market prices are made unstable that leads to gain their profits more.

\bibliographystyle{IEEEtran}
\bibliography{ref}

\end{document}